\documentclass[a4paper,11pt]{article}
\usepackage{amsmath, amssymb, graphicx, physics, bm}
\usepackage[a4paper, margin=1in]{geometry}
\usepackage{tikz}
\usetikzlibrary{arrows.meta,calc}
\usepackage{jheppub, appendix, mathrsfs} 

\title{\boldmath Stochastic Krylov Dynamics: Revisiting Operator Growth in Open Quantum Systems}

\author{Arpan Bhattacharyya$^{a}$}
\affiliation{$^{a}$Department of Physics, Indian Institute of Technology Gandhinagar, Gujarat-382355, India}
\author{S. Shajidul Haque$^{b,c}$, Jeff Murugan$^{b,c}$, Mpho Tladi$^{b}$ and Hendrik J.R. Van Zyl$^{b}$}
\affiliation{$^{b}$The Laboratory for Quantum Gravity \& Strings,\\
Department of Mathematics and Applied Mathematics,\\
University of Cape Town, Private Bag, Rondebosch, 7701,\\
South Africa}
\affiliation{$^{c}$National Institute of Theoretical and Computational Sciences (NITheCS),\\ 
Stellenbosch, 7604,\\ 
South Africa}

\emailAdd{abhattacharyya@iitgn.ac.in}
\emailAdd{jeff.murugan@uct.ac.za}
\emailAdd{hjrvanzyl@gmail.com}

\abstract{In closed quantum systems, Krylov complexity admits a geometric description: operator growth is equivalent to Hamiltonian flow in an emergent phase space whose structure is fixed by the Lanczos coefficients. We show that this picture survives, albeit in a fundamentally altered form, once the system is coupled to an environment. Using a Schwinger–Keldysh formulation of the full counting statistics of the Krylov position, we derive an effective action for operator growth under Lindblad dynamics. Even for the minimal case of pure dephasing, the phase-space dynamics ceases to be Hamiltonian: environmental coupling generates diffusion in the variable conjugate to Krylov depth, converting deterministic trajectories into stochastic ones. The hyperbolic mechanism underlying exponential complexity growth is therefore broadened and, beyond a parametrically controlled scale, destroyed. This identifies dissipation as a relevant perturbation of the chaotic Krylov fixed point and reveals operator growth in open systems as a problem of stochastic dynamics in an emergent phase space.
}

\begin{document}
\maketitle
\flushbottom

\section{Introduction}
\label{sec:intro}
Understanding how operators evolve and spread in quantum many-body systems has become a central theme in modern theoretical physics. This question lies at the heart of quantum chaos, thermalization, and information scrambling, and underpins phenomena ranging from the eigenstate thermalization hypothesis (ETH) to the fast scrambling conjecture and holographic duality \cite{Deutsch1991,Srednicki1994,Roberts2015,Shenker2014,Maldacena2016}. Traditionally, operator growth has been characterized using out-of-time-order correlators (OTOCs), which diagnose the sensitivity of quantum dynamics to perturbations and provide a notion of a quantum Lyapunov exponent \cite{Larkin1969,Shenker2014,Maldacena2016}. Complementary approaches based on spectral statistics and random matrix theory have further clarified the relationship between chaos, ergodicity, and universality in many-body systems \cite{Bohigas1984,Haake2010}.
More recently, Krylov complexity has emerged as a powerful and conceptually distinct diagnostic of operator growth \cite{Parker2019,Dymarsky:2019elm,Barbon2019,Barbon:2019wsy,Rabinovici:2020ryf,Avdoshkin_2020,Rabinovici2021}. Rather than focusing on specific correlation functions, Krylov complexity tracks how an operator spreads in the space generated by repeated commutation with the Hamiltonian. \textcolor{black}{This construction makes use of the computationally efficient Lanczos algorithm to measure operator growth and is closely tied to the structure of the Liouvillian.}

\noindent
Given a seed operator $O_0$, the Krylov construction generates an orthonormal sequence
\begin{eqnarray}
    \mathcal K = \text{span}\{O_0, \mathcal L O_0, \mathcal L^2 O_0, \dots\}\,,
\end{eqnarray}
where the Liouvillian superoperator is defined by
$\mathcal L O = [H,O]$.
Applying the Lanczos algorithm produces an orthonormal basis $\{|O_n\rangle\}$ in which the Liouvillian takes a tridiagonal form
\begin{eqnarray}
    \mathcal L |O_n\rangle = b_{n+1}|O_{n+1}\rangle + b_n|O_{n-1}\rangle\,,
\end{eqnarray}
with non-negative coefficients $b_n$, known as Lanczos coefficients. This representation is equivalent to a tight-binding Hamiltonian
\begin{eqnarray}
    \widehat H_{\textrm{TB}} = \sum_n b_{n+1}\Big(|n\rangle\langle n+1|+|n+1\rangle\langle n|\Big)\,,
\end{eqnarray}
acting on a semi-infinite chain, often referred to as the Krylov chain. Krylov complexity is then defined as the expectation value of the position operator on this chain,
\begin{eqnarray}
    \widehat n = \sum_n n |n\rangle\langle n|, \qquad K(t)=\langle \widehat n(t)\rangle,
\end{eqnarray}
and measures the average distance an operator has propagated in this Krylov space. This quantity captures operator growth in a coarse-grained but physically transparent way and has been shown to exhibit universal features across a wide range of systems, including exponential growth associated with chaotic dynamics and slower growth in integrable or constrained systems \cite{Parker2019,Barbon2019,Rabinovici2021} \footnote{For extension to the bosonic case readers are referred to \cite{Bhattacharyya:2023dhp}\,.}.\\

\noindent
A key recent conceptual advance is that Krylov dynamics admits a natural reformulation in terms of a real-time Schwinger–Keldysh path integral. In this approach, Krylov complexity is treated as an in–in observable generated by the full counting statistics
$Z(\chi,t)=\langle e^{i\chi \hat n(t)}\rangle$,
which plays the role of a dynamical generating functional. As shown in recent work \cite{Murugan:2026yyu}, this formulation maps operator growth onto classical Hamiltonian dynamics in an emergent phase space with coordinates $(n,p)$. The resulting effective Hamiltonian,
$H_{\rm eff}(n,p)=2b(n)\cos p$,
encodes the full information of the Lanczos coefficients and provides a geometric interpretation of operator growth. In particular, when $b(n)\sim \alpha n$, the phase-space flow develops hyperbolic fixed manifolds, and exponential growth of Krylov complexity emerges as a classical instability,
$K(t)\sim e^{2\alpha t}$.
This perspective not only unifies various observations about Krylov complexity but also enables a systematic classification of operator growth in terms of universality classes determined by the asymptotic scaling of $b(n)$.
Despite this progress though, the existing framework is restricted to closed quantum systems undergoing unitary evolution. In contrast, many experimentally relevant systems are inherently open; they interact with external environments, experience decoherence, and are subject to measurement backaction. Open quantum systems are typically described by Lindblad dynamics \cite{Lindblad1976,Gorini1976,Breuer:2007juk}, which introduces dissipation and noise at the level of the density matrix. These effects are known to qualitatively alter dynamical behavior, leading to phenomena such as decoherence, localization, and measurement-induced phase transitions \cite{Skinner2019,Chan2019}\footnote{Interested readers are referred to \cite{Chakrabarti:2025hsb} for a proposal of detecting measurement-induced phase transitions by observing dynamics of certain observable in Krylov space.}.\\

\noindent
The extension of Krylov complexity to such settings raises several conceptual challenges. The first, and most pressing, of which is that the Liouvillian governing operator evolution is no longer anti-Hermitian, and the Lanczos construction becomes non-unique or ill-conditioned. More fundamentally, it is not clear how the geometric phase-space picture of operator growth is modified in the presence of dissipation. The purpose of this work is to address these questions by extending the Schwinger–Keldysh formulation of Krylov complexity in \cite{Murugan:2026yyu} to open quantum systems. Working directly with the generating functional $Z(\chi,t)$, we derive a contour formulation of operator growth under Lindblad dynamics and identify the corresponding effective action in Krylov phase space. We show that environmental coupling generically converts the emergent Hamiltonian flow of the closed system into a stochastic dynamical system, with noise and dissipation encoded in the structure of the Keldysh action. This leads to a qualitatively new picture of operator growth in which hyperbolic instability competes with environmental fluctuations, providing a unified framework for studying complexity, scrambling, and dynamical phase transitions in open quantum systems.

\section{Linblad Dynamics and Open Systems}
\label{sec:Linblad}
Open quantum systems are generically described by completely positive, trace-preserving dynamics. In the Markovian limit, this evolution is governed by the Lindblad master equation \cite{Lindblad1976,Gorini1976,Breuer:2007juk}
\begin{equation}
    \dot\rho = -i[H,\rho] + \sum_\mu \left(L_\mu\rho L_\mu^\dagger - \frac12\left\{L_\mu^\dagger L_\mu ,\rho\right\}\right),
    \label{eq:Lindblad}
\end{equation}
where $H$ is the system Hamiltonian and the operators $L_\mu$ encode the coupling to the environment. These \textit{jump operators} describe processes such as dephasing, dissipation, particle loss, or continuous monitoring, and give rise to irreversible dynamics even in otherwise isolated systems.
While Eq.~\eqref{eq:Lindblad} is naturally formulated in the Schrödinger picture, operator growth is most naturally studied in the Heisenberg picture. In this case, observables evolve according to the adjoint Liouvillian
\begin{equation}
    \frac{d}{dt}O(t) = \mathcal L_{\rm   Lind}(O),
\end{equation}
with
\begin{equation}
    \mathcal L_{\rm Lind}(O) =
    i[H,O] + \sum_\mu \left( L_\mu^\dagger O L_\mu + \frac12\left\{L_\mu^\dagger L_\mu,O\right\} \right)\,.
    \label{Lindblad-Heisenberg}
\end{equation}
This equation generalizes the familiar Heisenberg evolution of closed systems by incorporating both coherent dynamics and dissipative processes. 

\subsection{Structure of the Lindbladian and loss of unitarity}
\label{subsec:Lind}
A crucial structural difference between closed and open systems lies in the properties of the generator of dynamics. For closed systems, the Liouvillian $\mathcal L=[H,\cdot]$ is anti-Hermitian with respect to the Hilbert–Schmidt inner product
\begin{equation}
    \langle A|B\rangle = \frac{1}{\mathcal N}\mathrm{Tr}(A^\dagger B)\,,
\end{equation}
and therefore generates unitary evolution in operator space. This property ensures that operator norms are preserved and that the dynamics can be interpreted as coherent motion in a Hilbert space of operators. In contrast, the Lindbladian $\mathcal L_{\rm Lind}$ is generically non-Hermitian and non-normal. As a result, the evolution in operator space is no longer unitary, operator norms are not conserved,
and the spectrum of $\mathcal L_{\rm Lind}$ can be complex.\\

\noindent
Physically, this reflects the presence of both decay and noise. The anti-commutator terms in Eq.~\eqref{Lindblad-Heisenberg} generate damping, while the $L_\mu^\dagger O L_\mu$ terms encode stochastic “re-injection” processes associated with environmental fluctuations. This combination leads to a qualitative change in the nature of operator dynamics. Instead of purely coherent spreading, operators can decay, localize, or exhibit nontrivial steady-state behavior.

\subsection{Operator growth in open systems}
In closed systems, operator growth is typically characterized by quantities such as OTOCs or Krylov complexity, which measure how an initially simple operator spreads over increasingly complex operator structures. In open systems, this notion must be generalized to account for the competing effects of coherent evolution and dissipation. Several new features arise in open systems:
\begin{itemize}
    \item \textbf{Competition between growth and decay:} The Hamiltonian part of $\mathcal L_{\rm Lind}$ drives operator spreading, while the dissipative terms tend to suppress it. Depending on the relative strength of these contributions, one can observe
	sustained growth, suppressed or slowed growth, or even saturation to a steady state. This competition is reminiscent of the interplay between scrambling and decoherence in a monitored quantum system \cite{Skinner2019}.
    \item \textbf{Absence of conserved operator norm:}
    In closed systems, the Hilbert-Schmidt norm of an operator is conserved, providing a natural notion of normalization for Krylov constructions. In open systems, this is no longer true since
    \begin{equation}
        \frac{d}{dt}\mathrm{Tr}(O^\dagger O)
        \neq 0\,.
    \end{equation}
    As a result, operator growth needs to be interpreted more carefully. We could, of course, normalize operators dynamically or focus only on normalized expectation values, but either choice introduces additional structure not present in the closed case.
    \item \textbf{Non-Hermitian Liouvillian and spectral structure:} The non-Hermitian nature of $\mathcal L_{\rm Lind}$ leads to a complex spectrum, with eigenvalues whose real parts correspond to decay rates and whose imaginary parts correspond to oscillatory dynamics. Consequently, the notion of a ``banded" or tridiagonal representation of the Liouvillian is no longer straightforward, and the connection between spectral properties and operator growth becomes more subtle. In particular, exponential growth associated with hyperbolic dynamics in closed systems may be replaced by transient growth followed by decay, or by entirely different scaling behavior.
    \item \textbf{Breakdown of the standard Krylov construction:}
    The Lanczos procedure relies crucially on the Hermiticity properties of the Liouvillian. When applied to $\mathcal L_{\rm Lind}$, the recursion coefficients may become complex, orthogonality of the Krylov basis is not guaranteed under the standard inner product, and the resulting tridiagonal representation need not correspond to a Hermitian tight-binding model.
    These complications can be circumvented by the use of various generalizations of the Lanczos algorithm to non-Hermitian operators, but they typically require bi-orthogonal bases or modified inner products \cite{Bhattacharya:2023zqt, Chakrabarti:2025hsb}, obscuring the physical interpretation of the Krylov chain. Consequently, while we can formally define a Krylov subspace for open systems,
    $\mathcal K = \text{span}\{O_0,\mathcal L_{\rm Lind} O_0,\mathcal L_{\rm Lind}^2 O_0,\dots\}$, 
     the resulting structure does not admit the same simple dynamical interpretation as in the closed case.
\end{itemize}

\subsection{Physical regimes of open-system operator growth}
\noindent
The interplay between coherent and dissipative dynamics leads to several qualitatively distinct regimes. In the \textit{coherent-dominated regime} weak dissipation allows operator growth to proceed similarly to the closed system, possibly with renormalized rates. On the other hand, in the \textit{dissipation-dominated regime}, strong environmental coupling suppresses operator spreading, leading to localization or rapid decay. In between these, there is a \textit{crossover regime} in which intermediate coupling produces nontrivial dynamics characterized by strong fluctuations and broad distributions of operator size. These regimes are not sharply separated in general and may depend sensitively on the choice of jump operators and the structure of the Hamiltonian. Understanding their universal features remains an open problem.\\

\noindent
To summarise then, while operator growth in closed systems admits a simple and geometrically transparent description in terms of Krylov dynamics, no comparable framework currently exists for open systems. The features of open systems discussed above suggest that operator growth in open systems should be viewed not as coherent propagation in a Hilbert space, but as a more general dynamical process involving both drift and fluctuations. It is the goal of this work to show that these features can be captured naturally by reformulating Krylov complexity in terms of a real-time generating functional. This approach 
leads to an emergent phase-space picture in which dissipation and noise appear as intrinsic components of the dynamics. In what follows, we adopt an effective description in which the environmental coupling is specified directly in Krylov space. While the underlying jump operators act in the microscopic Hilbert space, their projection onto the Krylov basis is generally complicated and nonlocal. Instead, we consider classes of jump operators that are simple in Krylov space, allowing for a controlled Schwinger–Keldysh formulation of open-system operator growth.

\section{Schwinger–Keldysh Generating Functional}
\noindent
The central object in our formulation \cite{Murugan:2026yyu} is the full counting statistics of the Krylov position operator,
\begin{equation}
    Z(\chi,t)
    = \left\langle e^{i\chi \hat n(t)} \right\rangle,
    \label{eq:FCS-def}
\end{equation}
where
\begin{equation}
    \hat n=\sum_{m\ge 0} m,|m\rangle\langle m|\,,
\end{equation}
is the position operator on the Krylov chain.  As in the closed-system setting, $Z(\chi,t)$ encodes the full probability distribution of the operator’s position in Krylov space.  Its derivatives generate the moments and cumulants of operator growth; in particular,
\begin{equation}
    K(t)=\langle \hat n(t)\rangle
    = \frac{\partial}{\partial(i\chi)}\,\ln Z(\chi,t)
    \Bigg|_{\chi=0}\,, \qquad
    \kappa_m(t) = \left. \frac{\partial^m}{\partial(i\chi)^m}\ln Z(\chi,t) \right|_{\chi=0}.
    \label{eq:cumulants}
\end{equation}
The virtue of the Schwinger–Keldysh formulation is that it provides a unified real-time generating functional for all of these quantities and makes their semiclassical and fluctuation structure manifest.

\subsection{Closed-system review}
Let us first recall the construction in the closed system discussed in our previous work \cite{Murugan:2026yyu}.  After Lanczos orthogonalization, the Heisenberg dynamics of the seed operator is equivalent to the Schrödinger evolution of a wavefunction on the semi-infinite Krylov chain,
\begin{equation}
    i\partial_t |\psi(t)\rangle = \widehat H_{TB} |\psi(t)\rangle\,,
\end{equation}
with tight-binding Hamiltonian
\begin{equation}
    \widehat H_{TB} = \sum_{m\ge 0} b_{m+1}\Big(|m\rangle\langle m+1|+|m+1\rangle\langle m|\Big)\,, 
    \label{eq:HTB}
\end{equation}
and initial state taken to be the seed,
\begin{equation}
    \rho_0 = |0\rangle\langle 0|\,.
    \label{eq:rho0}
\end{equation}
The quantity of interest is not a transition amplitude but an in–in expectation value. For this reason, the natural generating functional is defined on a closed time contour,
\begin{equation}
    Z[J_+,J_-] = \mathrm{Tr}\!\left( U_{J_+}(t,0)\,\rho_0\,U_{J_-}^\dagger(t,0) \right)\,,
    \label{eq:SK-gen-closed}
\end{equation}
where
\begin{equation}
    U_J(t,0) =
    \mathsf T \exp\!\left[ -i\int_0^t dt'\,\big(\widehat H_{TB}-J(t')\widehat n \big) \right]\,. 
    \label{eq:UJ-def}
\end{equation}
The two sources $J_+$ and $J_-$ live on the forward and backward branches of the contour, respectively. The physical expectation values are obtained by differentiating with respect to the quantum source $J_q(t)=J_+(t)-J_-(t)$. In particular,
\begin{equation}
    K(t) = \left. \frac{1}{i}\frac{\delta}{\delta J_q(t)} \ln Z[J_+,J_-] \right|_{J\pm=0}.
    \label{eq:K-from-Z}
\end{equation}
Eq.~\eqref{eq:SK-gen-closed} is already enough to emphasize the essential conceptual point that the Krylov problem is an instance of \textit{real-time full counting statistics}.  The counting field $\chi$ is simply the Fourier conjugate variable to the measured value of $\widehat n$, and $Z(\chi,t)$ is the characteristic function of the probability distribution of operator spreading. A convenient way to realize Eq.~\eqref{eq:FCS-def} in the contour formalism is to insert the counting field at the final time. We can do this either by a single insertion on one branch, or by a symmetric endpoint splitting on the two branches; the two prescriptions are equivalent up to cyclicity of the trace and differ only by convention. In either case, though, the contour construction reproduces
\begin{equation}
    Z(\chi,t) =
    \mathrm{Tr}\!\left(
    \rho_0\,e^{i\widehat H_{TB}t}\,e^{i\chi \widehat n}\,e^{-i\widehat H_{TB}t}
    \right) = \left\langle e^{i\chi \widehat n(t)}\right\rangle\,.
    \label{eq:FCS-Heisenberg}
\end{equation}
The next step is to rewrite Eq.~\eqref{eq:SK-gen-closed} as a path integral.  Time-slicing the evolution and inserting resolutions of identity in the Krylov basis,
\begin{equation}
    \mathbb I = \sum_{m\ge 0}|m\rangle\langle m|,
\end{equation}
on both branches converts the trace into a sum over discrete trajectories $m_k^\pm$ on the forward and backward contours.  The nontrivial step, described in detail in \cite{Murugan:2026yyu}, is to insert Fourier-complete momentum states
\begin{equation}
    |p\rangle = \sum_{m\ge 0} e^{ipm}|m\rangle\,,
    \qquad
    p\in(-\pi,\pi]\,,
    \label{eq:pstates}
\end{equation}
which exponentiates the nearest-neighbour hopping and produces a phase-space representation of the short-time kernel.  In the continuum limit, this gives the phase-space action
\begin{equation}
    S[n,p;J]
    = \int_0^t dt', \Big( p\,\dot n - H_{\rm eff}(n,p) + J(t')\,n
    \Big)\,,
    \label{eq:SnpJ}
\end{equation}
with the effective Hamiltonian
\begin{equation}
    H_{\rm eff}(n,p)=2\,b(n)\cos p\,.
    \label{eq:Heff}
\end{equation}
The closed-system generating functional, therefore, takes the form
\begin{equation}
    Z[J_+,J_-] = \int \mathscr D n_\pm\,\mathscr D p_\pm\; \exp\!\Big[iS[n_+,p_+;J_+]+ iS[n_-,p_-;J_-]\Big].
    \label{eq:closed-PI}
\end{equation}
Passing to Keldysh variables,
\begin{equation}
    n_c=\frac{n_+ + n_-}{2}\,,
    \qquad n_q=n_+-n_-\,, \qquad p_c=\frac{p_+ + p_-}{2}\,, \qquad p_q=p_+-p_-\,,
    \label{eq:Keldysh-vars}
\end{equation}
and expanding to leading order in the `quantum' fields gives the semiclassical action
\begin{equation}
    S_{\rm cl}
    = \int_0^t dt,\Big[ p_q\big(\dot n_c-\partial_p H_{\rm eff}(n_c,p_c)\big) + n_q\big(\dot p_c+\partial_n H_{\rm eff}(n_c,p_c)\big) \Big].
    \label{eq:Scl}
\end{equation}
Integration over $n_q$ and $p_q$ enforces Hamilton’s equations,
\begin{equation}
    \dot n_c=\partial_p H_{\rm eff}\,, \qquad
    \dot p_c=-\partial_n H_{\rm eff}\,,
    \label{eq:Hamilton-eqs}
\end{equation}
and exposes an emergent classical phase space with coordinates $(n,p)$.  It is precisely this structure that underlies the hyperbolic instability and universality classes discussed in the closed-system analysis.\\

\noindent
The closed-system construction already suggests how to proceed in the open case.  The essential observation is that the object of real interest is not the Lanczos recursion itself, but the contour generating functional $Z[J_+,J_-]$.  In the closed system, the latter may be rewritten as a doubled path integral whose semiclassical limit produces a classical Hamiltonian flow.  In the open system, we expect the same contour structure to remain valid, but with the action modified by the dissipative dynamics. This expectation is physically natural.  In an open system, the environment does two things simultaneously: it produces deterministic dissipative drift, and it generates fluctuations and decoherence. Both effects are intrinsically doubled-branch phenomena in that they arise from correlations between forward and backward histories, and therefore naturally enter as couplings between the + and - contours. Phrased differently, dissipation is not a correction to a single-branch Hamiltonian, but a genuinely contour-level effect.

\subsection{Open-system Schwinger–Keldysh formulation}
\noindent
To extend the closed-system construction to open quantum systems, we begin with Lindblad evolution\footnote{An alternative approach would be the bi-orthogonal approach to open systems.  In Appendix~(\ref{BiOrthogonalAppendix}) we show how this may be formulated in the Schwinger-Keldysh language.}, Eq.~\eqref{eq:Lindblad}.
As discussed earlier and in \cite{Murugan:2026yyu}, this evolution encodes both coherent dynamics and irreversible processes arising from coupling to an environment. Our goal is to incorporate these effects directly at the level of the real-time generating functional for operator growth. There are two conceptually equivalent routes to constructing the generating functional in the presence of dissipation:
\begin{enumerate}
    \item A microscopic route hinged on the Feynman-Vernon formalism \cite{Feynman:1963fq}, in which we would start from a system–bath Hamiltonian, integrate out the environmental degrees of freedom, and obtain a nonlocal influence functional \cite{Feynman:1963fq}. In the Markovian limit, this reduces to a local-in-time functional equivalent to Lindblad evolution.
    \item An effective route in which we construct the contour representation directly from the Lindblad superoperator, bypassing the explicit bath degrees of freedom.
\end{enumerate}
\noindent
For our purposes, the second route is more economical, but it is important to recognize that both approaches lead to the same structure. In particular, the dissipative contribution that appears in the contour action is precisely the Feynman–Vernon influence functional expressed in effective form. The generating functional generalizes Eq.~\eqref{eq:closed-PI} to
\begin{equation}
    Z[J_+,J_-]
    = \int \mathscr D(\text{fields})\,
    \exp\left( iS_{\rm sys}[+] - iS_{\rm sys}[-] + iS_{\rm diss}[+,-]
    \right)\,,
    \label{eq:open-SK-schematic}
\end{equation}
where $S_{\rm sys}$ is the coherent tight-binding action derived in the closed system,
and $S_{\rm diss}$ is a functional of both forward and backward trajectories induced by the Lindblad operators. The crucial new ingredient is $S_{\rm diss}[+,-]$, which encodes all environmental effects. Structurally, this term plays exactly the role of the Feynman–Vernon influence functional,
\begin{equation}
    \mathscr F[n_+,n_-] =
    \exp\!\left(iS_{\rm diss}[n_+,n_-]\right).
\end{equation}
In the original Feynman–Vernon formulation \cite{Feynman:1963fq}, the reduced density matrix can be written as a path integral over forward and backward trajectories,
\begin{equation}
    \rho_S =
    \int \mathscr D x_+ \mathscr D x_-\;
    e^{\,iS_S[x_+] - iS_S[x_-]}
    \mathscr F[x_+,x_-]
\end{equation}
where the influence functional $\mathscr F$ arises from integrating out the environment.
The key point is that $S_{\rm diss}[n_+,n_-]\;\equiv\;S_{\rm IF}[n_+,n_-]$, with the important distinction that, in our case, the ``coordinate" $n(t)$ is not a microscopic degree of freedom, but the emergent Krylov position describing operator growth. In other words, the influence functional acts directly in Krylov phase space. A defining feature of Eq.~\eqref{eq:open-SK-schematic} is that the action is no longer a simple difference of forward and backward contributions. Instead, the dissipative term introduces genuine inter-branch couplings. These couplings reflect the physical fact that decoherence suppresses interference between forward and backward histories.\\

\noindent
To understand their consequences, it is convenient to rewrite the action in Keldysh variables, Eq.~\eqref{eq:Keldysh-vars}.
The dissipative contribution then generically takes the form
\begin{equation}
    S_{\rm diss} =
    \int dt \left[
    \mathcal A(n_c,p_c)\,n_q +
    i\,\mathcal B(n_c,p_c)\,n_q^2
    +\cdots \right],
    \label{eq:generic-Sdiss}
\end{equation}
where $\mathcal A$ and $\mathcal B$ are model-dependent functions determined by the jump operators. The terms linear in $n_{q}$ modify the saddle-point equations and correspond to \textit{dissipative drift} in the emergent phase space, while the terms quadratic in $n_{q}$ encode \textit{fluctuations}. Upon integrating out the quantum fields, they generate stochastic noise in the effective dynamics. This structure is universal and is precisely the same decomposition that appears in the Feynman–Vernon theory of quantum Brownian motion, where the environment induces both friction and noise, related by fluctuation–dissipation relations.\\

\noindent
In the closed system, the action is linear in the quantum fields, and integrating them out enforces Hamilton’s equations. In the open system, the quadratic terms prevent this simplification.
Instead, integrating out the quantum fields leads to an effective weight of the form
\begin{eqnarray}
    \exp\!\left[
    -\int dt\, \frac{\big(\dot{n}_{c} - \partial_{p}H_{\mathrm{eff}}\big)^{2}}{\text{noise strength}} \;+\;\cdots
\right]\,,
\end{eqnarray}
which corresponds to stochastic dynamics for the classical variables. This modifies the emergent phase-space picture in two essential ways; the deterministic Hamiltonian flow is deformed by dissipative drift terms, and fluctuations introduce stochastic noise into the dynamics.

\section{Pure Dephasing}
\label{sec:dephasing}
To illustrate the general open-system construction in the simplest possible setting, we consider pure dephasing in the Krylov-position basis. This example is particularly transparent because the jump operator is diagonal in the natural basis of the emergent Krylov chain, so the effect of dissipation can be derived explicitly.

\subsection{Lindblad evolution in the Krylov basis}
\noindent
Consider a single Hermitian jump operator proportional to the Krylov position operator\footnote{We note here that this is an \textit{effective} jump operator acting in Krylov space. In Appendix~(\ref{sec:A}) we give an argument justifying this choice based on a microscopic-to-Krylov mapping.},
\begin{eqnarray}
    L=\sqrt{\kappa}\,\widehat n,
    \qquad \widehat n=\sum_{m\ge 0} m\,|m\rangle\langle m|\,.
\end{eqnarray}
Since $L$ is Hermitian, $L^\dagger=L = \sqrt{\kappa}\widehat n$ and $
L^\dagger L=\kappa \widehat n^2$. Substituting into the evolution equation for the density matrix, Eq.~\eqref{eq:Lindblad} gives
\begin{eqnarray}
    \dot{\rho} = -i[\widehat H_{\textrm{TB}},\rho] + \kappa\, \widehat n\,\rho\,\widehat n - \frac{\kappa}{2}\{\widehat n^2,\rho\}\,.
\end{eqnarray}

Equivalently, with a little algebra,
\begin{eqnarray}
    \dot{\rho} = -i[\widehat H_{\textrm{TB}},\rho] - \frac{\kappa}{2}[\widehat n,[\widehat n,\rho]]\,.
\end{eqnarray}
This is the standard pure-dephasing form. Let us now examine this equation in the Krylov basis. Writing $\rho = \sum_{m,m'} \rho_{mm'} |m\rangle\langle m'|$,
we have
\begin{eqnarray}
    (\widehat n\, \rho\, \widehat n)_{mm'} = m\,m' \,\rho_{mm'}\,,
    \qquad (\widehat n^2 \rho)_{mm'} = m^2 \rho_{mm'}\,, \qquad (\rho\, \widehat n^2)_{mm'} = m'^2 \rho_{mm'}\,.
\end{eqnarray}
The dissipative part therefore acts as
\begin{eqnarray}
    \left(\kappa\, \widehat n\,\rho\, \widehat n - \frac{\kappa}{2}\{\widehat n^2,\rho\}\right)_{mm'} &=& \kappa\, m\, m' \rho_{mm'} -\frac{\kappa}{2}(m^2+m'^2)\rho_{mm'}\nonumber\\
    &=& -\frac{\kappa}{2}(m-m')^2 \rho_{mm'}\,.
\end{eqnarray}
So the matrix elements obey
\begin{eqnarray}
    \dot{\rho}_{mm'} = -i(\hat H_{\textrm{TB}}\,\rho-\rho\, \hat H_{\textrm{TB}})_{mm'} -\frac{\kappa}{2}(m-m')^2 \rho_{mm'}\,.
\end{eqnarray}
This form makes the physics  explicit. The diagonal elements $m=m'$ are unaffected by the dissipator while off-diagonal elements $m\neq m'$ decay at a rate proportional to $(m-m')^2$.
Consequently, the environment suppresses coherence between different Krylov positions.\\

\noindent
Let's now translate this into the Schwinger–Keldysh language. In the closed system, the generating functional was written as a sum over forward and backward Krylov trajectories,
\begin{eqnarray}
    Z[J_+,J_-] = \sum_{\{m_k^+\},\{m_k^-\}} \mathcal{A}[m_k^+]\,\mathcal{A}[m_k^-]^*\, e^{\,i\sum_k \Delta t (J_k^+ m_k^+ - J_k^- m_k^-)}\,.
\end{eqnarray}
With Lindblad dephasing, each short time step acquires an extra factor coming from the dissipator. From the equation above, the density matrix element $\rho_{m_k^+,m_k^-}$ picks up
\begin{eqnarray}
    \rho_{m_k^+,m_k^-}(t+\Delta t) = \rho_{m_k^+,m_k^-}(t)\, \exp\!\left[ -\frac{\kappa}{2}\Delta t\,(m_k^+-m_k^-)^2
\right]\,,
\end{eqnarray}
to leading order in $\Delta t$.
Multiplying over time slices gives the total dissipative weight
\begin{eqnarray}
    \prod_k
\exp\!\left[
-\frac{\kappa}{2}\Delta t\,(m_k^+-m_k^-)^2
\right]
=
\exp\!\left[
-\frac{\kappa}{2}
\sum_k \Delta t\,(m_k^+-m_k^-)^2
\right].
\end{eqnarray}
Taking the continuum limit,
$m_k^\pm \to n_\pm(t)$, then gives
\begin{eqnarray}
    \exp\!\left[-\frac{\kappa}{2}\int_0^t dt'\,\big(n_+(t')-n_-(t')\big)^2 \right]\,.
\end{eqnarray}
This is the \textit{open-system influence functional} for pure dephasing and the full generating functional becomes
\begin{eqnarray}
    Z[J_+,J_-] &=& \int \mathscr{D}n_+\,\mathscr{D}p_+\, \mathscr{D}n_-\,\mathscr{D}p_-\; \exp\!\Big( iS[n_+,p_+;J_+] - iS[n_-,p_-;J_-] \Big)\nonumber\\
    &\times& \exp\!\left[ -\frac{\kappa}{2}\int_0^t dt'\,(n_+-n_-)^2 \right]\,.
\end{eqnarray}
This then is the precise open-system analogue of the closed-system phase-space path integral.

\subsection{Keldysh rotation and effective action}
\noindent
In terms of the `classical' and `quantum' Keldysh-variables, the dissipative factor in Eq.~\eqref{eq:Keldysh-vars} becomes simply
\begin{eqnarray}
    \exp\!\left[ -\frac{\kappa}{2}\int_0^t dt'\,n_q(t')^2 \right].
\end{eqnarray}
Equivalently, 
\begin{eqnarray}
    S_{\mathrm{diss}} = i\frac{\kappa}{2}\int_0^t dt'\,n_q^2\,,
\end{eqnarray}
resulting in the full Keldysh action
\begin{eqnarray}
    S_K = S_{\mathrm{cl}} + i\frac{\kappa}{2}\int_0^t dt'\,n_q(t')^2\,.
\end{eqnarray}
Now recall the coherent part. Expanding the closed-system action to first order in the quantum fields gives
\begin{eqnarray}
    S_{\mathrm{cl}} = \int_0^t dt\, \Big[p_q\big(\dot n_c-\partial_p H_{\mathrm{eff}}(n_c,p_c)\big) - n_q\big(\dot p_c+\partial_n H_{\mathrm{eff}}(n_c,p_c)\big) \Big]\,.
\end{eqnarray}
Finally, putting everything together gives
\begin{eqnarray}
    S_K = \int_0^t dt\, \Big[ p_q\big(\dot n_c-\partial_p H_{\mathrm{eff}}\big) - n_q\big(\dot p_c+\partial_n H_{\mathrm{eff}}\big) \Big] + i\frac{\kappa}{2}\int_0^t dt\,n_q^2\,,
\end{eqnarray}
which is the minimal open-system Keldysh action for Krylov dynamics under pure dephasing. From here we can now derive the equations of motion for the effective dynamics. Since $p_q$ appears linearly and without a quadratic term, its functional integral enforces
\begin{eqnarray}
    \dot n_c = \partial_p H_{\mathrm{eff}}(n_c,p_c)\,.
    \label{eq:n-equation}
\end{eqnarray}
So the equation for $n_c$ is unchanged by this minimal dephasing model. The $n_q$-dependent part of the action has the form
\begin{eqnarray}
    \int_0^t dt\,\left[ -n_q A(t)
    + i\frac{\kappa}{2}n_q^2 \right]\,,
\end{eqnarray}
where $A(t) \equiv \dot p_c + \partial_n H_{\mathrm{eff}}(n_c,p_c)$. The path integral over $n_q$ is therefore Gaussian,
\begin{eqnarray}
    \int \mathscr{D}n_q\, \exp\!\left[ -i\int dt\,n_q A(t) -\frac{\kappa}{2}\int dt\,n_q^2 \right].
\end{eqnarray}
Completing the square gives,
\begin{eqnarray}
    -\frac{\kappa}{2}n_q^2 - i n_q A = -\frac{\kappa}{2} \left( n_q + \frac{iA}{\kappa} \right)^2 -\frac{A^2}{2\kappa}\,,
\end{eqnarray}
so that
\begin{eqnarray}
    \int \mathscr{D}n_q\, e^{\,iS[n_q]} \propto \exp\!\left[-\frac{1}{2\kappa}\int_0^t dt\,A(t)^2 \right].
\end{eqnarray}
Hence the effective weight after integrating out $n_q$ is
\begin{eqnarray}
    \exp\!\left[ -\frac{1}{2\kappa}\int_0^t dt\, \big(\dot p_c+\partial_n H_{\mathrm{eff}}(n_c,p_c)\big)^2 \right]\,,
    \label{eq:effective-weight}
\end{eqnarray}
which is the standard Martin–Siggia–Rose/Onsager–Machlup form of a stochastic process \cite{MartinSiggiaRose1973,OnsagerMachlup1953}. To understand and interpret this, let's reduce the problem to something a bit more familiar, discretise the time variable $t\to t_i$ and define $A_i \equiv A(t_i)$. Then the weight $\displaystyle\to \exp\!\left[ -\frac{1}{2\kappa}\sum_i A_i^2 \right]$.
Now introduce auxiliary variables $\eta_i$ and write the identity,
\begin{eqnarray}
    \exp\!\left[-\frac{1}{2\kappa}A_i^2\right] = \int \frac{d\eta_i}{\sqrt{2\pi\kappa}} \exp\!\left[ -\frac{1}{2\kappa}\eta_i^2 \right] \delta(\eta_i - A_i)\,.
\end{eqnarray}
Multiplying over all $i$ results in
\begin{eqnarray}
    \exp\!\left[-\frac{1}{2\kappa}\sum_i A_i^2\right] = \int \prod_i \frac{d\eta_i}{\sqrt{2\pi\kappa}} \exp\!\left[ -\frac{1}{2\kappa}\sum_i \eta_i^2 \right] \prod_i \delta(\eta_i - A_i)\,.
\end{eqnarray}
Finally, taking the continuum limit, this becomes
\begin{eqnarray}
    \exp\!\left[ -\frac{1}{2\kappa}\int dt\,A(t)^2 \right] = \int \mathscr{D}\eta\; \exp\!\left[ -\frac{1}{2\kappa}\int dt\,\eta(t)^2 \right] \delta\!\big[\eta(t)-A(t)\big]\,.
\end{eqnarray}
This identity is a statement that $\eta(t)$ is distributed with Gaussian weight
\begin{eqnarray}
    P[\eta] \propto \exp\!\left[-\frac{1}{2\kappa}\int dt\,\eta^2\right]\,,
\end{eqnarray}
which in turn implies $\langle \eta(t)\eta(t')\rangle = \kappa\,\delta(t-t')$. The delta functional then enforces
$\eta(t)=A(t)$. So averaging over $\eta$ with this constraint is exactly equivalent to weighting configurations by
\begin{eqnarray}
    \exp\!\left[-\frac{1}{2\kappa}\int dt\,A(t)^2\right]\,.
\end{eqnarray}
In our case, $A(t) = \dot p + \partial_n H_{\mathrm{eff}}$, with weight mentioned in Eq.~\eqref{eq:effective-weight}. Using the identity above, this is equivalent to
\begin{eqnarray}
    \int \mathscr{D}\eta\; \exp\!\left[ -\frac{1}{2\kappa}\int dt\,\eta^2 \right] \delta\!\big[\dot p + \partial_n H - \eta(t)\big]\,.
\end{eqnarray}
The delta functional enforces $\dot p + \partial_n H_{\text{eff}} = \eta(t)$
or
\begin{eqnarray}
    \dot p = -\partial_n H_{\text{eff}} + \eta(t)\,,
\end{eqnarray}
with $\langle \eta(t)\eta(t')\rangle = \kappa\,\delta(t-t')$ and which, together with Eq.~\eqref{eq:n-equation} constitute \textit{Langevin equations} for the stochastic effective dynamical system governing operator growth in the open system. The key point here is a Gaussian weight in the square of an equation $\sim e^{-\int (\text{equation})^2}$
is equivalent to saying that the equation is not exactly satisfied, but instead that it is satisfied up to Gaussian fluctuations. Notice that noise appears in the equation conjugate to the variable whose quantum field is squared.

\section{Operator Growth in Chaotic Open Systems}
\label{sec:linear-lanczos}
\noindent
To build on the intuition developed in the last section for pure dephasing, let's specialize further to the case of linear Lanczos coefficients $b(n)=\alpha\, n$. This is a useful case study because the closed-system flow is exactly the hyperbolic fixed-point dynamics, and the open-system deformation can be analyzed analytically, as we will show below. In this case, the effective Hamiltonian $H_{\rm eff}(n,p)=2\alpha\, n \cos p$, and the stochastic equations are
\begin{eqnarray}
    \dot n&=&\partial_p H_{\rm eff}=-2\alpha n \sin p\,,\nonumber\\
    \dot p&=&-\partial_n H_{\rm eff}+\eta(t)=-2\alpha \cos p+\eta(t)\,,
\end{eqnarray}
with white noise $\langle \eta(t)\eta(t')\rangle=\kappa\,\delta(t-t')$. This is essentially a minimal model for noisy hyperbolic flow. When $\kappa=0$, the phase portrait is determined by the classical dynamical system
\begin{eqnarray}
    \dot n&=&-2\alpha\, n \sin p\,,\nonumber\\
    \dot p&=&-2\alpha\, \cos p\,.
\end{eqnarray}
The special manifolds can be found at $p=\pm \frac{\pi}{2}$. At
$p=-\pi/2$, we have
\begin{eqnarray}
    \dot p=0\,,
    \qquad
    \dot n=2\alpha n\,,
\end{eqnarray}
so $n(t)=n_0 e^{2\alpha t}$.
This is the unstable hyperbolic direction responsible for exponential Krylov growth. At
$p=+\pi/2$, we have
\begin{eqnarray}
    \dot p=0\,,
    \qquad
    \dot n=-2\alpha n\,,
\end{eqnarray}
resulting in a contracting manifold. In the closed-system growth mechanism is clear; if the trajectory is steered toward $p=-\pi/2$, it grows exponentially in $n$.\\

\noindent
To analyze the open system, we expand around the unstable branch as
\begin{eqnarray}
    p(t)=-\frac{\pi}{2}+\delta p(t)\,, \qquad |\delta p|\ll 1\,.
\end{eqnarray}
Then
$\sin p = -\cos \delta p \simeq -1+\frac{\delta p^2}{2}$, $\cos p = \sin \delta p \simeq \delta p$, and the stochastic equations become
\begin{eqnarray}
    \dot n &=& -2\alpha n\, \sin p
    \simeq 2\alpha n\left(1-\frac{\delta p^2}{2}\right)\,,\nonumber\\
    \\
    \dot{\delta p} &=& -2\alpha\, \delta p+\eta(t)\,.\nonumber
    \label{eq:O-U-process}
\end{eqnarray}
This is the key simplification; near the unstable manifold, the noisy hyperbolic flow reduces to exponential growth in $n$ driven by a restoring-but-noisy process for the angular variable $\delta p$. The stochastic differential equation $\dot{\delta p}=-2\alpha \delta p+\eta(t)$ is readily integrated to give
\begin{eqnarray}
    \delta p(t) = \delta p(0)e^{-2\alpha t} + \int_0^t ds\, e^{-2\alpha (t-s)}\eta(s)\,.
    \label{eq:stochastic-solution}
\end{eqnarray}
There are several points about this solution that deserve some unpacking:
\begin{itemize}
    \item Taking expectation values on both sides of Eq.~\eqref{eq:stochastic-solution} and using the fact that the white noise function has zero mean, we find the mean value
    \begin{eqnarray}
        \langle \delta p(t)\rangle = \delta p(0)e^{-2\alpha t}\,,
    \end{eqnarray}
    from which it is clear that deterministic drift pulls the trajectory toward the unstable manifold.
    \item Similarly, its variance 
    \begin{eqnarray}
        \mathrm{Var}\big[\delta p(t)\big] &=& \left\langle \left( \int_0^t ds\, e^{-2\alpha (t-s)}\eta(s) \right)^2 \right\rangle\,,\nonumber\\ 
        &=& \kappa \int_0^t ds\, e^{-4\alpha (t-s)} = \frac{\kappa}{4\alpha}\left(1-e^{-4\alpha t}\right)\,,
    \end{eqnarray}
    tells us that at late times,
    \begin{eqnarray}
        \langle \delta p^2\rangle_{\rm st}= \frac{\kappa}{4\alpha}\,,
    \end{eqnarray}
    so the environment broadens the unstable manifold into a strip of width $\delta p_{\rm rms}\sim \sqrt{\kappa/4\alpha}$. This is a concrete manifestation of noisy hyperbolic flow.
    \item The two-time correlator is
    \begin{eqnarray}
        \langle \delta p(t)\delta p(t')\rangle &=& \Bigg\langle \Bigg[\delta p(0)e^{-2\alpha t} + \int_0^t ds\, e^{-2\alpha (t-s)}\eta(s) \Bigg] \Bigg[ \delta p(0)e^{-2\alpha t'} + \int_0^{t'} ds'\, e^{-2\alpha (t'-s')}\eta(s') \Bigg] \Bigg\rangle\nonumber\\
        &=& \delta p(0)^2 e^{-2\alpha (t+t')} + \int_0^t ds \int_0^{t'} ds'\, e^{-2\alpha (t-s)}e^{-2\alpha (t'-s')} \langle \eta(s)\eta(s')\rangle\,,
    \end{eqnarray}
    assuming the initial condition is deterministic and uncorrelated with the noise. Using
    $\langle \eta(s)\eta(s')\rangle=\kappa\,\delta(s-s')$, we get
    \begin{eqnarray}
        \langle \delta p(t)\delta p(t')\rangle &=& \delta p(0)^2 e^{-2\alpha (t+t')} + \kappa \int_0^{\min(t,t')} ds\, e^{-2\alpha (t-s)}e^{-2\alpha (t'-s)}\nonumber\\
        &=&  \delta p(0)^2 e^{-2\alpha (t+t')} + \kappa e^{-2\alpha (t+t')} \int_0^{\min(t,t')} ds\, e^{4\alpha s}\nonumber\\
        &=& \delta p(0)^2 e^{-2\alpha (t+t')} + \frac{\kappa}{4\alpha} \left( e^{-2\alpha |t-t'|} - e^{-2\alpha (t+t')} \right).
    \end{eqnarray}
    This expression gives the \textit{exact two-time correlator}. At long times, the transient pieces proportional to $e^{-2\alpha(t+t')}$ disappear and the correlator takes the time-translation invariant Ornstein–Uhlenbeck form
    \begin{eqnarray}
        \langle \delta p(t)\delta p(t')\rangle_{\rm st} =
        \frac{\kappa}{4\alpha}\,e^{-2\alpha |t-t'|}\,.
    \end{eqnarray}
    From this, we can immediately read off the variance $\langle \delta p^2\rangle_{\rm st}=\kappa/4\alpha$ and correlation time $\tau_p=1/2\alpha$ and infer that $\alpha$ controls not only the hyperbolic growth scale in $n$, but also how quickly angular fluctuations relax back toward the unstable manifold.
\end{itemize}
The above analysis makes it clear that the open system exhibits two competing tendencies. Noise injects fluctuations with strength $\kappa$, while deterministic drift restores the system toward the unstable manifold with rate $2\alpha$. If $\kappa\ll \alpha$, the strip is narrow, and trajectories remain well aligned with the unstable direction.
If $\kappa\gtrsim \alpha$, the strip becomes broad, and the notion of a sharply resolved unstable manifold is lost. The exponential decay $e^{-2\alpha |t-t'|}$
means that the angular wandering is not free Brownian motion. It is colored by the restoring drift and fluctuations persist only over a time scale $1/(2\alpha)$. Physically, this means the environment continually kicks the system away from the unstable manifold, but the deterministic hyperbolic flow pulls it back.
In other words, the open-system growth is not simply a case of ``noise on top of chaos". It is a dynamically balanced competition between instability and restoration.

\subsection{Consequences for operator growth}
Recall that expanding near the unstable branch produces
\begin{eqnarray}
    \dot n = 2\alpha n\left(1-\frac{\delta p^2}{2}+\cdots\right)\,.
    \label{eq:n-growth-rate}
\end{eqnarray}
This means that the growth rate is modulated by $\delta p^2(t)$. The two-time correlator above tells us how these rate fluctuations are correlated in time. Specifically, because $\delta p(t)$ is correlated over the scale $1/(2\alpha)$, the suppression of growth is not instantaneous and uncorrelated; it comes in temporally coherent bursts. Consequently, we should expect pure dephasing open systems to exhibit intermittent trajectory-to-trajectory behavior, and enhanced higher cumulants and large-deviation effects. In this sense, the two-time correlator is the first direct signature that the open system should exhibit a nontrivial trajectory ensemble, not just a renormalized mean exponent.\\

\noindent
To make the previous point more concrete, let's define the local growth rate by
\begin{eqnarray}
    \lambda(t)\equiv \frac{d}{dt}\ln n(t)\,.
\end{eqnarray}
Then, near the unstable manifold,
$\lambda(t)\simeq 2\alpha-\alpha\,\delta p(t)^2$ and fluctuations in $\lambda(t)$ are governed by correlations of $\delta p^2$. Since $\delta p$ is Gaussian, these can be computed by Wick’s theorem to get
\begin{eqnarray}
    \langle \delta p^2(t)\delta p^2(t')\rangle_{\rm st} &=& 2\langle \delta p(t)\delta p(t')\rangle_{\rm st}^2 + \langle \delta p^2\rangle_{\rm st}^2\nonumber\\
    &=& 2\left(\frac{\kappa}{4\alpha}\right)^2 e^{-4\alpha |t-t'|} + \left(\frac{\kappa}{4\alpha}\right)^2\,,
\end{eqnarray}
where we have used our previous expressions for the two-time correlator and variance in the last equality. The key point here is that the connected correlator of $\delta p^2$ decays on the shorter time scale $\frac{1}{4\alpha}$. This, in turn, means the growth-rate fluctuations themselves have finite memory and are temporally correlated, precisely the stochastic mechanism underlying the fluctuation-dominated crossover discussed earlier.

\subsection{Effective Renormalized Growth Exponent}
In the preceding section, we used a simple variance estimate to claim that fluctuations in the growth rate are controlled by correlations in $\delta p^2$, leading to $\lambda(t)\simeq 2\alpha-\alpha\,\delta p(t)^2$. Let's be more systematic. Starting from the expansion near the unstable manifold
\begin{eqnarray}
    \frac{\dot n}{n} = 2\alpha\left(1-\frac{\delta p^2}{2}\right)\,,
\end{eqnarray}
we get
\begin{eqnarray}
    \frac{d}{dt}\ln n(t) = 2\alpha - \alpha\,\delta p(t)^2\,.
\end{eqnarray}
Integrating this and taking expectation values on both sides gives
\begin{eqnarray}
    \left\langle \ln \frac{n(t)}{n_0} \right\rangle &=& 2\alpha t - \alpha \int_0^t ds\,\langle \delta p(s)^2\rangle\nonumber\\
    &=& 2\alpha t - \alpha \int_0^t ds \left[ \delta p(0)^2 e^{-4\alpha s} + \frac{\kappa}{4\alpha}\left(1-e^{-4\alpha s}\right)\right]\,,
\end{eqnarray}
where we have used our previously derived expression for $\langle \delta p(s)^2\rangle$. The integrals in this exact-to-quadratic-order expression are both elementary and easily computed to give,
\begin{eqnarray}
    \left\langle \ln \frac{n(t)}{n_0} \right\rangle = \left(2\alpha-\frac{\kappa}{4}\right)t - \frac{\delta p(0)^2}{4}\left(1-e^{-4\alpha t}\right) + \frac{\kappa}{16\alpha}\left(1-e^{-4\alpha t}\right).
\end{eqnarray}
At a late time,
\begin{eqnarray}
    \left\langle \ln n(t)\right\rangle \sim \left(2\alpha-\frac{\kappa}{4}\right)t + \mathcal O(1)\,,
\end{eqnarray}
so the \textit{typical growth exponent} is
\begin{eqnarray}
    \lambda_{\rm typ}=2\alpha-\frac{\kappa}{4}\,.
\end{eqnarray}
To see why this is the correct typical exponent, note that it is this exponent that governs
\begin{eqnarray}
    \lim_{t\to\infty}\frac{1}{t}\langle \ln n(t)\rangle\,,
\end{eqnarray}
making it the natural Lyapunov-like growth rate for a multiplicative stochastic process. It characterizes the behavior of a typical trajectory. This is not the same as 
\begin{eqnarray}
    \lim_{t\to\infty}\frac{1}{t}\ln \langle n(t)\rangle\,,
\end{eqnarray}
which can be dominated by rare trajectories and is therefore not the correct quantity for the ``typical" growth of operator complexity in a noisy environment. This then justifies our claim that \textit{pure dephasing renormalizes the typical hyperbolic growth exponent from $2\alpha\text{ to }2\alpha-\kappa/4$}.\\

\noindent
To go beyond the mean, note that 
\begin{eqnarray}
    \left\langle \lambda(t)\right\rangle = 2\alpha-\alpha\,\left\langle \delta p(t)^2\right\rangle\,,
\end{eqnarray}
so in the stationary regime,
$\langle \lambda\rangle_{\rm st} =
2\alpha-\frac{\kappa}{4}$. This can, in turn, be used to compute the connected two-time correlator of $\lambda$ as
\begin{eqnarray}
    C_\lambda(t-t') = \langle \lambda(t)\lambda(t')\rangle_c = \alpha^2 \left\langle \delta p(t)^2 \delta p(t')^2\right\rangle_c\,.
\end{eqnarray}
Again, because $\delta p$ is Gaussian, Wick’s theorem gives
\begin{eqnarray}
    \left\langle \delta p(t)^2 \delta p(t')^2\right\rangle_c = 2\left\langle \delta p(t)\delta p(t')\right\rangle^2.
\end{eqnarray}
Using the stationary correlator 
$\langle \delta p(t)\delta p(t')\rangle_{\rm st} = \frac{\kappa}{4\alpha}e^{-2\alpha |t-t'|}$, it follows, with a little algebra, that
\begin{eqnarray}
    C_\lambda(\tau) = \frac{\kappa^2}{8}\,e^{-4\alpha |\tau|}\,.
\end{eqnarray}
Physically, this tells us that while the mean growth rate is reduced linearly in $\kappa$, fluctuations of the growth rate are of order $\kappa^2$, and they decorrelate on the time scale
$\tau_\lambda=\frac{1}{4\alpha}$.
So growth occurs in temporally correlated bursts rather than via uncorrelated white-noise modulation.\\

\noindent
All in all, this gives a sharper picture of the noisy hyperbolic phase. In the closed system, the unstable manifold $p=-\pi/2$ produces a fixed local growth rate $\lambda = 2\alpha$. In the open system, the trajectory wanders around that manifold. The mean wandering lowers the typical rate, and the finite-time correlations of that wandering induce fluctuations in the local growth rate. In other words, the open-system dynamics is not just ``exponential growth with a smaller exponent." It is a genuinely stochastic hyperbolic process. The two key quantities in this process are the stationary angular width $\delta p_{\rm rms}=\sqrt{\kappa/(4\alpha)}$,
which determines how far trajectories spread away from the unstable direction, and the growth-rate correlator $\displaystyle  C_\lambda(\tau)=\frac{\kappa^2}{8}e^{-4\alpha|\tau|}$, which determines how persistent the local scrambling fluctuations are.
This explains the three regimes discussed earlier:
\begin{itemize}
    \item $\kappa\ll \alpha$: the phase space contains a narrow stochastic tube corresponding to weakly renormalized chaos;
    \item $\kappa\sim \alpha$: in which the tube broadens, inducing large intermittent fluctuations;
    \item $\kappa\gg \alpha$: in which the unstable manifold is washed out, resulting in no typical exponential growth.
\end{itemize}

\section{Schwinger–Keldysh Formulation of a Non-Hermitian Krylov Chain}
\label{sec:non-herm-chain}
A central goal of our work in this article is to develop a unified description of operator growth in open quantum systems. In the previous sections, we have constructed a Schwinger–Keldysh formulation directly from the Krylov chain of a closed system and showed how environmental effects enter as influence functionals that modify the emergent phase-space dynamics.\\

\noindent
There is, however, an alternative and complementary approach that begins by projecting the full Lindblad dynamics onto the Krylov basis of the closed system. This leads to an effective \textit{non-Hermitian} tight-binding model for Krylov amplitudes, which has been used to analyze the suppression of operator growth and the emergence of edge-localized modes in \cite{Liu:2022god, Bhattacharya:2023yec, Chakrabarti:2025hsb}. In this section, we will re-derive this effective non-Hermitian Krylov Hamiltonian from first principles, construct its Schwinger-Keldysh formulation, and interpret its long-time behavior in the phase-space language developed above. This establishes a direct bridge between the spectral, non-Hermitian perspective and our trajectory-based Schwinger-Keldysh framework, and clarifies how dissipation reorganizes operator growth at a dynamical level.\\

\subsection{Dissipative Projection}
Recall from Eq.~\eqref{Lindblad-Heisenberg} that the full Lindbladian takes the form
\begin{eqnarray}
    \mathcal L_{\rm Lind} = \mathcal L_0 + \mathcal D\,.
\end{eqnarray}
We would like to understand how the dissipative term $\mathcal D$ acts in the Krylov basis defined by $\mathcal L_0$. To this end, projecting onto this basis, we define
\begin{eqnarray}
    D_{mn}\equiv -\,\langle O_m | \mathcal D | O_n \rangle\,.
\end{eqnarray}
This quantity measures how dissipation mixes and attenuates Krylov modes. For Hermitian jump operators $L_\mu=L_\mu^\dagger$, the dissipator simplifies to
\begin{eqnarray}
    \mathcal D(O) =-\frac{1}{2}\sum_\mu [L_\mu,[L_\mu,O]]\,,
\end{eqnarray}
which yields the explicit representation
\begin{eqnarray}
    D_{mn} = \frac{1}{2}\sum_\mu \langle [L_\mu,O_m]\,|\, [L_\mu,O_n]\rangle\,.
\end{eqnarray}
This expression makes the structure of dissipation transparent. In particular, the diagonal elements
\begin{eqnarray}
    D_{nn} = \frac{1}{2}\sum_\mu \|[L_\mu,O_n]\|^2 \ge 0\,,
\end{eqnarray}
are manifestly non-negative and quantify how strongly the operator $|O_n\rangle$ fails to commute with the environment. In other words, they provide a natural measure of how susceptible a given Krylov mode is to decoherence.\\

\noindent
Projecting the full evolution onto the Krylov basis and defining amplitudes $\phi_n(t)=\langle O_n|O(t)\rangle$, gives
\begin{eqnarray}
    i\partial_t \phi_n = -b_{n+1}\phi_{n+1} - b_n \phi_{n-1} -i\gamma\sum_m D_{nm}\phi_m\,,
    \label{eq:non-hermitian-hopping}
\end{eqnarray}
where $\gamma$ is the overall dissipation strength inherited from the microscopic Lindblad equation. This is equivalent to Schrödinger evolution with an effective non-Hermitian Hamiltonian
\begin{eqnarray}
    \widehat H_{\rm NH} = \widehat H_{\textrm{TB}} - i\gamma\,\widehat D,
    \label{eq:non-hermitian-hamiltonian}
\end{eqnarray}
where $\widehat D=\sum_{m,n} D_{mn}|m\rangle\langle n|$. Thus, dissipation appears as an imaginary potential in Krylov space, which attenuates the amplitudes of different Krylov modes. In general, the matrix $D_{mn}$ is not diagonal. However, for local dissipators and chaotic dynamics, one expects a form of emergent locality in Krylov space in the sense that operators at a given Krylov depth have similar complexity and therefore similar overlap with the dissipative channels. In this regime, the dominant contribution comes from the diagonal elements,
\begin{eqnarray}
    \widehat D \approx d(\widehat n), \qquad d(\widehat n)=\sum_n d_n |n\rangle\langle n|\,.
\end{eqnarray}
Empirically and theoretically, it is known that $d_n$ grows approximately linearly with Krylov depth before saturating at a finite-size scale, $d_n \sim \beta_d + \alpha_d n$. Under this approximation, the effective Hamiltonian reduces to $\widehat H_{\rm NH} \approx \widehat H_{\textrm{TB}}
- i\,d(\widehat n)$. This tells us that dissipation acts as an imaginary potential that penalizes operator complexity.

\subsection{Schwinger-Keldysh formulation of the non-Hermitian Krylov chain}
Having derived the effective non-Hermitian Hamiltonian, we are now in a position to construct a real-time generating functional for operator growth directly in Krylov space. Conceptually, this step parallels the construction carried out earlier for closed systems, but with an important distinction: the evolution is no longer unitary, and expectation values must therefore be defined with an appropriate normalization.\\

\noindent
The projection procedure of the previous subsection shows that dissipation enters as an imaginary potential that depends on Krylov depth. From the perspective of dynamical observables, this implies that trajectories in Krylov space are not only governed by the coherent hopping encoded in $\widehat H_{\textrm{TB}}$, but are also weighted by a decay factor that penalizes excursions to large $n$. The role of the Schwinger–Keldysh formalism is to make this weighting precise and to embed it into a contour framework that allows us to compute normalized correlation functions and full counting statistics. In particular, the non-conservation of norm implies that the natural object is not a simple expectation value, but a normalized generating functional, which plays the role of a dynamical partition function for Krylov trajectories. This will allow us to reinterpret the non-Hermitian evolution in terms of a path integral over trajectories, in which dissipation appears as an absorptive functional that competes with coherent operator growth.\\

\noindent
The non-unitary evolution generated by the effective Hamiltonian $\widehat H_{\rm NH} = \widehat H_{\textrm{TB}} - i\gamma d(\widehat n)$
requires a careful definition of dynamical observables. In particular, expectation values must be computed with respect to a time-dependent normalization, since the norm of the evolved state is not conserved. This leads naturally to the normalized generating functional
\begin{eqnarray}
    \mathcal G(\chi,t) = \frac{ \langle 0|e^{i\hat H_{\rm NH}^\dagger t}\,e^{i\chi \hat n}\,e^{-i\hat H_{\rm NH}t}|0\rangle}{\langle 0|e^{i\hat H_{\rm NH}^\dagger t}e^{-i\hat H_{\rm NH}t}|0\rangle}\,.
    \label{eq:norm-gen-fun}
\end{eqnarray}
This object plays the role of a dynamical partition function for Krylov trajectories, and generates moments of the Krylov position operator,
\begin{eqnarray}
    K(t) = \left.\frac{\partial}{\partial(i\chi)}\ln \mathcal G(\chi,t)\right|_{\chi=0}.
    \label{eq:non-herm-krylov}
\end{eqnarray}
\noindent
To cast this into a Schwinger–Keldysh form, we introduce branch-dependent sources $J_\pm(t)$ coupled to $\widehat n$, and define forward and backward evolution operators
\begin{eqnarray}
    U_+(t,0) &=& \mathsf T \exp\!\left[-i\int_0^t dt'\,(\widehat H_{\rm NH}-J_+(t')\widehat n)\right],\nonumber\\
    U_-^\dagger(t,0) &=& \widetilde{\mathsf T}\exp\!\left[ +i\int_0^t dt'\,(\widehat H_{\rm NH}^\dagger-J_-(t')\widehat n) \right]\,.
\end{eqnarray}
The normalized generating functional then becomes
\begin{eqnarray}
   \mathcal G[J_+,J_-] =\frac{ \langle 0|U_-^\dagger(t,0)\,U_+(t,0)|0\rangle}{\langle 0|U_-^\dagger(t,0)\,U_+(t,0)|0\rangle\big|_{J_\pm=0}}\,. 
\end{eqnarray}
This is the exact Schwinger-Keldysh representation of the non-Hermitian Krylov dynamics. The crucial structural difference from the closed-system case in \cite{Murugan:2026yyu} is that the backward branch evolves with $\widehat H_{\rm NH}^\dagger$, reflecting the non-unitarity of the dynamics.\\

\noindent
Following the same steps as in the closed-system construction \cite{Murugan:2026yyu}, we can express the generating functional as a path integral over trajectories $(n_\pm(t),p_\pm(t))$ on the forward and backward branches. The coherent part of the action is
\begin{eqnarray}
    S[n,p;J] = \int_0^t dt'\,\Big(p\,\dot n - H_{\rm eff}(n,p) + J(t')\,n\Big)\,,
\end{eqnarray}
where again $H_{\rm eff}(n,p)=2b(n)\cos p$. The non-Hermitian contribution produces an additional weight. Using
$\widehat H_{\rm NH} = \widehat H_{\textrm{TB}} - i\gamma\, d(\widehat n)$, and 
$\widehat H_{\rm NH}^\dagger = \widehat H_{\textrm{TB}} + i\gamma\, d(\widehat n)$, we find that the forward and backward branches contribute identically to the real exponential damping, yielding
\begin{eqnarray}
    \exp\!\left[-\gamma\int_0^t dt'\,\big(d(n_+)+d(n_-)\big) \right]\,.
\end{eqnarray}
Consequently, the full generating functional takes the form
\begin{eqnarray}
    \mathcal G[J_+,J_-] = \frac{\int \mathcal D n_\pm\,\mathcal D p_\pm\; \exp\!\left[iS[n_+,p_+;J_+] - iS[n_-,p_-;J_-]-\gamma\int_0^t dt'\,\big(d(n_+)+d(n_-)\big) \right].}{\int \mathcal D n_\pm\,\mathcal D p_\pm\; \exp\!\left[iS[n_+,p_+;0] - iS[n_-,p_-;0]-\gamma\int_0^t dt'\,\big(d(n_+)+d(n_-)\big) \right]}\,.\nonumber\\
\end{eqnarray}
This expression shows that the non-Hermitian dynamics can be interpreted as a reweighting of closed-system trajectories by an exponential decay factor depending on their Krylov depth.\\

\subsection{Effective action}
To compute the associated effective action, we again follow the construction in \cite{Murugan:2026yyu} and 
introduce the classical and quantum fields
\begin{eqnarray}
    n_c=\frac{n_+ + n_-}{2}\,, \quad n_q=n_+-n_-\,, \qquad p_c=\frac{p_+ + p_-}{2}\,, \quad p_q=p_+-p_-
\end{eqnarray}
and expand the action in powers of the quantum fields. The coherent part yields
\begin{eqnarray}
    S_{\rm cl} = \int_0^t dt\, \Big[p_q(\dot n_c - \partial_p H_{\rm eff}) - n_q(\dot p_c + \partial_n H_{\rm eff})\Big].
\end{eqnarray}
For the dissipative term, we expand
\begin{eqnarray}
    d(n_+) + d(n_-) = 2d(n_c) + \frac{1}{4}d''(n_c)\,n_q^2 + O(n_q^4)\,,
\end{eqnarray}
and substitute into the exponent to get
\begin{eqnarray}
    iS_{\rm eff} = iS_{\rm cl} -2\gamma\int_0^t dt\,d(n_c) -\frac{\gamma}{4}\int_0^t dt\,d''(n_c)\,n_q^2 +\cdots.
\end{eqnarray}
This effective action has a clear structure:
\begin{itemize}
    \item The term $\displaystyle -2\gamma\int d(n_c)$ acts as a trajectory-dependent survival weight, suppressing paths that explore large Krylov depth, and
    \item The quadratic term $\propto d''(n_c)\,n_q^2$ would generate fluctuations, but it vanishes when $d(n)$ is linear.
\end{itemize}
In the regime $d(n)\simeq \beta_d + \alpha_d n$, which is relevant for local dissipators and chaotic dynamics, $d''(n)=0$, so the effective action reduces to
\begin{eqnarray}
    iS_{\rm eff} = iS_{\rm cl}-2\gamma\int_0^t dt\,(\beta_d + \alpha_d n_c)\,.
\end{eqnarray}
In other words, the Schwinger-Keldysh theory becomes deterministic but non-unitary, so classical trajectories are unchanged, but their contribution to the generating functional is exponentially suppressed according to their time-integrated Krylov depth.

\subsection{Saddle-point dynamics and long-time asymptotics}
The Schwinger–Keldysh formulation derived above provides a path-integral representation of the non-Hermitian Krylov dynamics in terms of trajectories in the emergent phase space $(n(t),p(t))$. We now analyze this theory at the saddle-point level and use it to extract the long-time behavior of Krylov complexity. Toward this end, we start from the effective Keldysh action in the form
\begin{eqnarray}
    iS_{\rm eff} = i\int_0^t dt\,\Big[p_q(\dot n_c - \partial_p H_{\rm eff})- n_q(\dot p_c + \partial_n H_{\rm eff})\Big]-2\gamma\int_0^t dt\,d(n_c)+\cdots.
\end{eqnarray}
Since the quantum fields $n_q$ and $p_q$ appear linearly (for linear $d(n)$), they act as Lagrange multipliers enforcing the classical equations of motion. Varying with respect to $p_q$ and $n_q$ therefore gives
\begin{eqnarray}
    \dot n_c = \partial_p H_{\rm eff}(n_c,p_c)\,, \qquad \dot p_c = -\partial_n H_{\rm eff}(n_c,p_c)\,. \label{eq1}
\end{eqnarray}
At saddle-point level, the dynamics of the classical variables $(n_c,p_c)$ is identical to that of the closed system. In particular, dissipation does not modify the equations of motion themselves. Instead, its effect is entirely encoded in the real exponential weight
\begin{eqnarray}
    \exp\!\left[-2\gamma\int_0^t dt\,d(n_c(t))\right]\,,
\end{eqnarray}
which multiplies each trajectory in the path integral. This leads to a simple but important reinterpretation of the dynamics; the Schwinger-Keldysh theory describes an ensemble of classical Hamiltonian trajectories, with each trajectory weighted by a survival probability determined by its time-integrated Krylov depth. In this sense, the dynamics obeys a kind-of ``principle of least decay", rather than a modification of the classical equations of motion.\\

\noindent
The behavior of the system is  controlled by a competition between
\textit{coherent operator growth}, governed by the Hamiltonian flow generated by $H_{\rm eff}$, and
\textit{dissipative suppression}, governed by the functional $\displaystyle\int dt\,d(n_c(t))$.
To make this competition explicit, consider the regime relevant for chaotic systems, in which
$b(n)\simeq \alpha_b n$, and
$d(n)\simeq \beta_d + \alpha_d n$.
In the absence of dissipation, the classical equations of motion exhibit a hyperbolic instability, leading to exponential growth along the unstable manifold,
$n_c(t) \sim n_0 e^{2\alpha_b t}$.
Substituting this into the dissipative weight gives
\begin{eqnarray}
    \int_0^t dt'\,d(n_c(t')) \;\simeq\; \beta_d t + \alpha_d \int_0^t dt'\,n_c(t')\,.
\end{eqnarray}
The first term produces ordinary exponential decay. The second term dominates at long times. Using the exponential growth of $n_c(t)$, we obtain
\begin{eqnarray}
   \int_0^t dt'\,n_c(t') \sim\frac{n_0}{2\alpha_b}e^{2\alpha_b t}\,. 
\end{eqnarray}
Therefore the trajectory weight behaves as
\begin{eqnarray}
    \exp\!\left[-2\gamma\int_0^t dt'\,d(n_c(t'))\right]\;\sim\; \exp\!\left[-\frac{\gamma\alpha_d n_0}{\alpha_b}\,e^{2\alpha_b t}\right],
\end{eqnarray}
and is \textit{super-exponentially} suppressed.\\

\noindent
This result has a crucial consequence. Although the classical dynamics still allows trajectories with exponentially growing $n_c(t)$, such trajectories are overwhelmingly suppressed in the path integral. The suppression is so strong that they make a negligible contribution to the generating functional at long times. Instead, the dominant contributions come from trajectories that minimize the functional $\displaystyle\int_0^t dt\,d(n_c(t))$. Since $d(n)$ is an increasing function of $n$, this selects trajectories that remain as close as possible to the edge of the Krylov chain, $n\simeq 0$. In other words, long-lived trajectories are those localized near the edge of Krylov space.
This provides a direct phase-space interpretation of the edge-localized eigenmodes that arise in the spectral analysis of the non-Hermitian Hamiltonian. Both descriptions reflect the fact that complex operators decay rapidly, while simple operators are comparatively long-lived.\\

\noindent
Since the generating functional is normalized, the expectation value of $\widehat n(t)$ is dominated at late times by these least-decaying trajectories. As a result, Krylov complexity no longer grows indefinitely, but instead saturates to a finite value set by the typical Krylov depth of the surviving trajectories. Specifically, 
\begin{eqnarray}
    K(t)\xrightarrow[t\to\infty]{} K_\infty \sim O(1)\,,
\end{eqnarray}
up to possible weak dependence on system size through the saturation scale of $d(n)$. The long-time saturation value $K_\infty$ may be estimated from the dominant edge-localized mode of the non-Hermitian chain,
\begin{eqnarray}
    K_\infty \approx \frac{\sum_n n\,|\phi^{(\star)}_n|^2}{\sum_n |\phi^{(\star)}_n|^2}\,,
\end{eqnarray}
with the dominant right eigenvector,
$\phi_n^{(\star)} \propto e^{-n/\xi}$. Then, on a finite Krylov chain of dimension\footnote{Here $D_K$ denotes the largest Krylov index generated by the Lanczos procedure. Since the basis is labeled by $n=0,1,\dots,D_K$, the Krylov space has dimension $D_K+1$. We will use $D_K$ to refer to the maximal Krylov depth, rather than the dimension itself.} $D_K+1$, this gives the exact expression
\begin{eqnarray}
    K_\infty(D_K,\xi) \approx \frac{\sum_{n=0}^{D_K} n\, e^{-2n/\xi}}{\sum_{n=0}^{D_K} e^{-2n/\xi}} = \frac{q\big(1-(D_K+1)q^{D_K} + D_K q^{D_K+1}\big)} {(1-q)(1-q^{D_K+1})}\,,
\end{eqnarray}
with $q\equiv e^{-2/\xi}$.
This expression makes it clear that the large-$\xi$ behavior depends on the order of limits. If we first take $D_K\to\infty$, then 
\begin{eqnarray}
   K_\infty(\xi) &=&
   \frac{\sum_{n=0}^{\infty} n q^n}{\sum_{n=0}^{\infty} q^n} = \frac{e^{-2/\xi}}{1-e^{-2/\xi}}=\frac{1}{e^{2/\xi}-1}\,,
\end{eqnarray}
using standard expressions for geometric sums. Then expanding this expression for large $\xi$ gives
\begin{eqnarray}
    K_\infty(\xi) &=& \frac{1}{e^{2/\xi}-1} = \frac{\xi}{2}\left(1+\frac{1}{\xi}+O(\xi^{-2})\right)^{-1}\nonumber\\
    &=& \frac{\xi}{2}-\frac12+O(\xi^{-1})\,.
\end{eqnarray}
On the other hand, if we take the limit $\xi\to\infty$ at fixed $D_K$ then, since
$q=e^{-2/\xi}\to 1$, we have
$q^n \to 1$ for each fixed $n\le D_K$ and the numerator and denominator become
\begin{eqnarray}
    \sum_{n=0}^{D_K} n\,q^n \to \sum_{n=0}^{D_K} n =  \frac{D_K(D_K+1)}{2}\,,
\end{eqnarray}
and 
\begin{eqnarray}
    \sum_{n=0}^{D_K} q^n \to \sum_{n=0}^{D_K} 1 =  D_K+1\,,
\end{eqnarray}
respectively. Therefore
\begin{eqnarray}
    K_\infty(D_K,\xi)\to \frac{\frac{D_K(D_K+1)}{2}}{D_K+1} = \frac{D_K}{2}.
\end{eqnarray}
In the former case, balancing the coherent growth scale $b(n)\sim \alpha_b n$ against the dissipative potential $\gamma d(n)\sim \gamma\alpha_d n$ yields $\xi\sim \alpha_b/(\gamma\alpha_d)$, and hence
\begin{eqnarray} \label{eq:saturation}
    K_\infty \sim \frac{\alpha_b}{\gamma\alpha_d}\,,
\end{eqnarray}
up to a nonuniversal numerical prefactor. This estimate makes explicit that the saturation value decreases with increasing dissipation strength and increases with the coherent operator-growth rate. The finite system
saturation value is \(\min(D_K/2,\; \xi/2)\) (up to some small corrections). The overall picture that emerges is that the underlying classical dynamics retains its hyperbolic structure and continues to generate exponentially growing trajectories. However, the non-Hermitian weight strongly suppresses such trajectories. The long-time dynamics is therefore governed not by the most rapidly growing trajectories, but by the most slowly decaying ones. In this way, dissipation does not eliminate the mechanism of operator growth at the level of equations of motion, but instead reorganizes the statistical ensemble of trajectories that contribute to physical observables.
\subsection{Scrambling vs Dissipation}
The preceding analysis reveals a clean physical picture: operator growth in an open quantum system is governed by a race between two competing processes. On one hand, the coherent Hamiltonian dynamics drives exponential growth of the Krylov complexity along the unstable manifold, attempting to scramble information across the system. On the other hand, dissipation, whether in the form of stochastic noise (Sections (\ref{sec:dephasing}) -(\ref{sec:linear-lanczos})) or an absorptive potential (Section (\ref{sec:non-herm-chain})) acts to suppress, slow, or destroy this growth.\\

\noindent
This competition can be quantified by comparing two characteristic time scales. To this end, we define the bare scrambling time as the time required for a closed system to
spread an operator to the end of the Krylov chain, \textit{i.e.}, to reach the Krylov dimension \(D_K\):
\[
t_{\mathrm{scr}}^{\mathrm{(bare)}} = \frac{1}{2\alpha_b}\ln\left(\frac{D_K}{n_0}\right)\,,
\]
where $n_0$ is the initial Krylov position and $\alpha_b$ is the asymptotic slope 
of the Lanczos coefficients, \(b(n) \sim \alpha_b n\).
Scrambling, the operator reaching the end of the Krylov chain, occurs if the
bare scrambling time is shorter than the dissipation time controlled by the localization length $\xi$:
\begin{eqnarray}
    \frac{1}{2\alpha_b}\ln\left(\frac{D_K}{n_0}\right) \;<\; \frac{1}{2\alpha_b}\ln\left(\frac{\xi}{n_0}\right) \quad\Longrightarrow\quad D_K < \xi\,.
\end{eqnarray}
The system scrambles if the Krylov dimension \(D_K\)
is smaller than the localization length \(\xi\). Conversely, if \(D \gg \xi\), the state localizes before reaching the boundary and scrambling is suppressed.
When the Krylov dimension \(D_K\) is smaller than the localization length
\(\xi\) (i.e., \(D_K \lesssim \xi\)), the operator reaches the boundary of
Krylov space before dissipation can kill it; scrambling proceeds essentially
as in the closed system. Conversely, when \(D_K \gg \xi\), the environment
absorbs the operator before it can fully scramble. Note that, here $D_k$ is finite.\\

\noindent
The same argument applies to the stochastic regime of Sections (\ref{sec:dephasing}) and (\ref{sec:linear-lanczos}), though the 
competition takes a different form. There, noise does not absorb trajectories but instead reduces the typical growth exponent to $\lambda_{\mathrm{typ}} = 2\alpha - 
\kappa/4$. Scrambling requires $\lambda_{\mathrm{typ}} > 0$, i.e.
\begin{equation}
\kappa \;<\; 8\alpha .
\end{equation}
In both cases, the open system exhibits a dynamical phase transition controlled by the relative strength of coherent growth and environmental coupling. The unifying message is that operator growth in open quantum systems is not simply a deformed version of the closed-system problem, but rather a genuine race whose 
outcome determines whether information scrambling survives or is destroyed by dissipation.

\section{Conclusions and Outlook}
In this work, we have developed a unified framework for understanding operator growth in open quantum systems by extending the Schwinger–Keldysh formulation of Krylov complexity in \cite{Murugan:2026yyu} to Lindblad dynamics. Starting from the full counting statistics of the Krylov position operator, we constructed a real-time generating functional that provides a 
description of operator growth beyond unitary evolution. This formulation naturally leads to an emergent phase-space picture in which the dynamics of operator spreading is governed by an effective action defined on a doubled contour. Our central result is that environmental coupling qualitatively modifies the geometric picture of Krylov dynamics. In closed systems, operator growth is governed by Hamiltonian flow in an emergent phase space, with exponential complexity growth arising from hyperbolic instability. In open systems, this structure survives but is fundamentally altered since dissipation introduces both drift and fluctuations at the level of the Schwinger–Keldysh action, converting deterministic trajectories into stochastic ones.\\

\noindent
This mechanism is made explicit in Section (\ref{sec:dephasing}) and (\ref{sec:linear-lanczos}), where we analyzed the minimal case of pure dephasing. There, the Keldysh action acquires a quadratic term in the quantum field, leading to Langevin dynamics in the conjugate variable $p$. The resulting picture is that of noisy hyperbolic flow, in which exponential growth persists but is renormalized and broadened by fluctuations. The instability responsible for operator growth is no longer sharply defined; instead, it is smeared into a stochastic tube whose width is set by the noise strength. This leads to a reduction of the typical growth exponent, enhanced trajectory-to-trajectory fluctuations, and nontrivial temporal correlations in the growth rate. Operator growth in this regime is therefore intrinsically dynamical, governed by a competition between instability and noise.\\

\noindent
In contrast, Section (\ref{sec:non-herm-chain}) presents a complementary perspective based on projecting Lindblad dynamics onto the Krylov basis, yielding an effective non-Hermitian tight-binding model. In this formulation, the Schwinger–Keldysh action remains deterministic at leading order, and dissipation enters not through stochasticity but through an absorptive weight that suppresses trajectories that explore large Krylov depth. The resulting dynamics is governed by a principle of least decay; while hyperbolic trajectories continue to exist at the level of equations of motion, they are exponentially suppressed in the path integral. Consequently, long-time behavior is dominated by edge-localized trajectories, and Krylov complexity saturates to a finite value set by the spatial profile of the least-decaying mode.\\

\noindent
These two pictures are not contradictory but rather represent different limits of the same underlying Schwinger–Keldysh structure. One emphasizes fluctuation-driven dynamics, while the other emphasizes spectral selection and survival. Their coexistence highlights that operator growth in open systems is not a simple deformation of the closed-system problem, but a genuinely richer dynamical phenomenon. This competition is captured by a simple race condition. In the non-Hermitian regime, scrambling occurs when the Krylov dimension
\(D_K\) is smaller than the localization length \(\xi\) (i.e., \(D_K \lesssim \xi\)). When this holds, the operator reaches the boundary of Krylov space before
dissipation kills it; otherwise, dissipation wins and the operator never fully scrambles. In the stochastic regime, the race is instead between the bare growth
exponent \(2\alpha\) and the noise strength \(\kappa\), with scrambling requiring \(\kappa < 8\alpha\). This unifying perspective reveals a dynamical phase
transition whose outcome is determined by the relative strength of coherent growth and dissipation.\\

\noindent
The framework developed here raises several concrete and, we believe, high-impact questions that go beyond incremental extensions. Among these are:
\begin{itemize}
    \item \textbf{Universality classes of open-system operator growth:} In closed systems, the asymptotic scaling of Lanczos coefficients defines universality classes of operator growth. Our results suggest that dissipation acts as a relevant perturbation of the chaotic fixed point, but in two sharply distinct ways: as stochastic noise (Sections \ref{sec:dephasing}-\ref{sec:linear-lanczos}) and as an absorptive potential (Section \ref{sec:non-herm-chain}). A key open question is therefore:
    \textit{What are the universality classes of operator growth in open systems?} Do these two mechanisms define distinct fixed points, or are they different representations of a single universality class? Is there an RG flow in Krylov space in which noise and decay correspond to different directions? Answering this would promote Krylov complexity from a diagnostic to a classification tool for non-equilibrium quantum dynamics.
    \item \textbf{Thermal baths and fluctuation–dissipation structure:} A natural and important extension of the present framework is to environments that satisfy a fluctuation–dissipation relation, \textit{i.e.} genuine thermal baths. In contrast to the dephasing models considered here—where noise and dissipation can be introduced independently, a thermal environment ties them together through the Kubo–Martin–Schwinger (KMS) condition. In the Schwinger–Keldysh formulation, this implies that the drift (retarded) and noise (Keldysh) components of the action are not independent, but are constrained by temperature. As a result, the emergent phase-space dynamics will generically contain both friction-like terms and stochastic fluctuations with amplitudes fixed by the bath temperature, leading to a thermally constrained competition between instability, damping, and noise. From the perspective of Krylov dynamics, this opens up qualitatively new behavior. Rather than simply broadening hyperbolic growth (as in pure dephasing) or suppressing it through trajectory selection (as in the non-Hermitian case), a thermal bath is expected to drive the system toward a steady-state distribution in Krylov space, determined by a balance between coherent growth and dissipation. This suggests that Krylov complexity may acquire a thermodynamic interpretation, with temperature controlling the effective occupation of Krylov modes and potentially defining new universality classes of operator growth. Understanding this interplay, and whether it leads to sharp crossovers or phase transitions, would connect the present framework directly to quantum thermodynamics and open-system chaos.

    \item \textbf{Extensions to interacting and spatially structured systems:} The present framework reduces operator growth to dynamics on a one-dimensional Krylov chain, effectively capturing complexity through a single coordinate n, but in doing so suppresses additional structure intrinsic to many-body systems, such as spatial locality and conservation laws. A natural extension is therefore to develop a Schwinger–Keldysh description in which operator growth carries both spatial and Krylov degrees of freedom, leading to a coupled phase-space dynamics that can capture features such as front propagation, anisotropic spreading, and modified light-cone structures in the presence of dissipation. Incorporating conserved quantities would further enrich this picture by introducing hydrodynamic constraints that can compete with or slow down scrambling, potentially giving rise to sub-ballistic or diffusive regimes of complexity growth. Such generalizations would not only deepen the conceptual understanding of operator dynamics in open systems, but also enable direct comparison with experimental and numerical studies of interacting quantum matter, where spatial structure and conservation laws play a central role.
\end{itemize}

\noindent
In summary, we have shown that operator growth in open quantum systems can be understood as a problem of dynamical evolution in an emergent phase space, enriched by the interplay of instability, noise, and dissipation. The Schwinger–Keldysh framework developed here provides a natural language for this problem and lays the groundwork for a systematic theory of complexity in open quantum matter.

\section*{Acknowledgements}
JM would like to acknowledge the organisers of the HEPCAT 2026 research retreat for a stimulating and productive environment where some of this work was done. SSH, JM  and HJRVZ are supported in part by the ``Quantum Technologies for Sustainable Development" grant
from the National Institute for Theoretical and Computational Sciences of South Africa
(NITheCS). AB is supported by the Core Research Grant (CRG/2023/ 001120) by the Department of Science and Technology (DST) and Anusandhan National Research Foundation (ANRF), India. AB also acknowledges the associateship program of the Indian Academy of Sciences (IASc), Bengaluru, and acknowledges support from the Indian Institute of Technology Gandhinagar and a generous donor through the Singheswari and Ram Krishna Jha Chair. AB also acknowledges various discussions with Saptaswa Ghosh related to the Schwinger-Keldysh method and for current collaboration on a project based on this approach.   

\appendix

\section{A Bi-Lanczos Approach}

\label{BiOrthogonalAppendix}
In the main text, we are interested in understanding the complexity of open systems in the ``canonical" formulation, making use of the Krylov basis and complexity operator that follows from the system Hamiltonian.  An alternative approach to treating open systems is in terms of the bi-orthogonal basis.  In this case, it is possible to introduce two bases $|p_n\rangle$ and $|q_n\rangle$ from a bi-Lanczos algorithm \cite{bilanczos,Bhattacharya:2023zqt, NSSrivatsa:2023pby,bhattacharjee:2023uwx, Carolan:2024wov, Liu:2024stj,Chakrabarti:2025hsb,Baggioli:2025knt,Bhattacharyya:2025lsc}.  Briefly, as a starting point, we take some reference state $|\phi_r\rangle$
and $|p_0\rangle = |\phi_r\rangle$, $|q_0\rangle = |\phi_r\rangle$.  We then generate the following set of states iteratively
\begin{eqnarray}
|p_{n+1}) & = & \mathcal{L} |p_n) - \frac{(q_n| \mathcal{L} |p_n) }{ (q_n | p_n)  } |p_n) - \frac{(q_{n-1}| \mathcal{L} |p_n) }{ (q_{n-1} | p_{n-1})  } |p_{n-1})\,,    \nonumber \\
|q_{n+1}) & = & \mathcal{L}^\dag |q_n) - \frac{(p_n| \mathcal{L}^\dag |q_n) }{ (p_n | q_n)  } |q_n) - \frac{(p_{n-1}| \mathcal{L}^\dag |q_n) }{ (p_{n-1} | q_{n-1})  } |q_{n-1}) \,.   \nonumber
\end{eqnarray}
Note that e.g. we have 
$$   (q_0| \mathcal{ L } |p_2)  =    \overline{  (p_2| \mathcal{ L }^\dag |q_0)}  = \overline{(p_2| q_1   )} + \overline{(p_2| q_0   )}  \frac{\overline{(p_2| \mathcal{ L }^\dag |q_0) } }{   \overline{  (p_0| q_0)    }}   = (q_1| p_2)  +    \frac{ (q_0 | \mathcal{L}| p_2)}{(q_0|p_0)}  (q_0|p_2)  = 0  $$
so that no further projections are needed to be subtracted in order to ensure that 
$$ ( q_m | p_n   ) = \delta_{mn} (q_n| p_n) \,.$$
We thus introduce the ``normalised" states\footnote{At this stage we have not yet declared that these \textbf{are} the normalised state, so we are still working with the usual Hilbert space inner product.  If we declare these as the normalised states, then this should be viewed as a deformation of the Hilbert space inner product.} 
$$ |p_n\rangle \equiv \frac{1}{\sqrt{(q_n|p_n)}} |p_n) \ \ \ ; \ \ \ |q_n\rangle \equiv \frac{1}{\sqrt{(q_n|p_n)}} |q_n)$$
Note that these states are not necessarily normalised with respect to the Hilbert space inner product \textit{i.e.} $\langle p_n | p_n\rangle \neq 1$.  
The Liouvillian is given as \footnote{Note, that we could also have represented the Liouvillian in the $|p_n\rangle\langle p_m|$ basis, but it would not be tri-diagonal in that case.}  
$$
\mathcal{L} = \sum_{n} c_{n+1} |p_{n+1}\rangle \langle q_n| + b_{n+1} |p_{n}\rangle \langle q_{n+1}| + a_n |p_{n}\rangle \langle q_n|\,. $$
The conjugate Liouvillian is given by 
$$
\mathcal{L}^\dag = \sum_{n} b_{n+1}^* |q_{n+1}\rangle \langle p_n| + c_{n+1}^* |q_{n}\rangle \langle p_{n+1}| + a_n^* |q_{n}\rangle \langle p_n|\,. $$
The (bi-orthogonal) Krylov complexity is defined in analogy to the usual case.  The time-evolved reference ket can be expanded as
$$ |p_0(t)\rangle = e^{- i t \mathcal{L}} |p_0\rangle = \sum_{n \geq 0} \phi_n(t) |p_n\rangle $$
and the time-evolved reference bra as 
$$ \langle q_0(t)| = \langle q_0 | e^{ i t \mathcal{L}} = \sum_{n \geq 0} (\psi_n(t))^* \langle q_n|\,.$$   
The expansion coefficients may be obtained as $ \phi_n(t)  = \langle q_n| e^{- i t \mathcal{L}} |p_0\rangle$ and  \hbox{$ (\psi_n(t))^*  = \langle q_0 | e^{ i t \mathcal{L}} |p_n\rangle$}.  
The generating function for the (bi-orthogonal) Krylov complexity is then defined as 
$$  Z = \sum_{n\geq 0} \phi_n(t) (\psi_n(t))^*  e^{i \chi n}$$
which we may rewrite as
\begin{equation}
Z = \sum_n \langle q_n| e^{- i t \mathcal{L}} |p_0\rangle \langle q_0 | e^{ i t \mathcal{L}} |p_n\rangle  e^{i \chi n}\,.
\end{equation}
Crucially, the identity operator may be represented as 
$$ I = \sum_{n} |q_n\rangle \langle p_n|\,.$$
As usual, we can now compute the above by a time-slicing procedure with insertions of resolutions of the identity after each evolution by a small time.  For example, we thus have
\begin{equation}
\langle q_n| e^{- i t \mathcal{L}} |p_0\rangle = \sum_{n_1, n_2, \cdots, n_N} \prod_{ j=0  }^N \langle q_{j+1}| e^{- i \Delta t \mathcal{L}} |p_j\rangle
\end{equation}
where $q_{N+1} = q_n$ and $\Delta t = \frac{t}{N+1}$.  To proceed from here, we may make use of the family of kets
\begin{equation}
|P\rangle = \sum_{n}    e^{i P n} |p_n\rangle
\end{equation}
and bras 
\begin{equation}
\langle P | = \sum_{n}    e^{-i P n} \langle q_n|\,.
\end{equation}
These satisfy the two useful properties.  First, we have a resolution of the identity through
$$   I = \frac{1}{2\pi} \int_{-\pi}^\pi |P\rangle\langle P|  $$
and we have a simple expression for the overlaps
$$ \langle q_n |P\rangle = e^{i P n} \ \ \ ; \ \ \  \langle P | p_n \rangle = e^{-i P n}\,. $$
Using this we compute 
$$ \langle q_{j+1}| e^{- i \Delta t \mathcal{L}} |p_j\rangle =  \frac{1}{2\pi} \int_{-\pi}^\pi dP_j \langle q_{j+1}|P_j\rangle \langle P_j| e^{- i \Delta t \mathcal{L}} |p_j\rangle \approx \int_{-\pi}^\pi dP_j e^{- i t H_{\textrm{eff}}(  j, P_j  )}\,. $$
In the continuum limit, using the midpoint prescription for evaluating the matrix elements, we have 
$$ H_{\textrm{eff}}(n,p) = b(n+\frac{1}{2})e^{i p} + a(n) + c(n-\frac{1}{2})e^{- i p}\,. $$
Returning now to the generating function, this is an in-out observable that we may compute with a Schwinger-Keldysh path integral.  On the forward branch, we time-evolve with the Hamiltonian $H_{\textrm{eff}}$ while on the backward branch, we evolve with its conjugate $H_{\textrm{eff}}^\dag$.  Putting it all together, this gives rise to the generating function
\begin{equation}
Z_{SK}\left[ J_{+}, J_{-} \right] = \int \mathcal{D} n_{+} \mathcal{D} p_{+} \mathcal{D} n_{-}  \mathcal{D} p_{-} \textnormal{exp}\left( i S_{SK}'\left[ n_{+}, p_{+}; \chi_{+}   \right] - i S_{SK}\left[ n_{-}, p_{-}; \chi_{-}   \right]   \right)
\end{equation}
where 
\begin{eqnarray}
S_{SK}[n, p; \chi] = \int_{0}^t dt' \left( p \dot{n} - H_{\textrm{eff}}(n, p) + \chi(t') n    \right)\,,   \nonumber \\
S_{SK}'[n, p; \chi] = \int_{0}^t dt' \left( p \dot{n} - H_{\textrm{eff}}^\dag(n, p) + \chi(t') n    \right)\,.
\end{eqnarray}
The boundary conditions are (like in the closed system case) imposed by the initial state and trace conditions, giving $n_{+}(0) = n_{-}(0) = 0$ and $n_{+}(t) = n_{-}(t)$.  To compute Krylov complexity, we need to take derivatives w.r.t. the multipliers $\chi_{\pm}$ that provide insertions on both the forward and backward branches of the evolution.  Explicitly, we have 
$$ K(t) = \left. \frac{1}{2i} \left( \frac{\delta}{\delta J_{+}(t)} +   \frac{\delta}{\delta J_{-}(t)}   \right)  \ln Z_{SK}[J_{+}, J_{-}] \right|_{J_{\pm} = 0}\,.$$

\section{Microscopic-to-Krylov mapping of jump operators}
\label{sec:A}
In the main text we have adopted an effective description in which the environmental coupling is specified directly in Krylov space. While the underlying jump operators act in the microscopic Hilbert space, their projection onto the Krylov basis is generally complicated and nonlocal, motivating us to consider classes of jump operators that are simple in Krylov space, allowing for a controlled Schwinger–Keldysh formulation of open-system operator growth. This viewpoint 
decouples operator-growth physics from microscopic details and lets us classify universality classes of open-system scrambling and derive phase diagrams analytically. Nevertheless, it is worth a short digression to justify the form of the effective jump operator.\\

\noindent
To start, consider a 1D spin-$\tfrac12$ chain with Hamiltonian
$H = \sum_j h_{j,j+1}$. A generic nonintegrable nearest-neighbor model such as
\begin{eqnarray}
    H= J\sum_j \sigma_j^z\sigma_{j+1}^z+g\sum_j \sigma_j^x+h\sum_j \sigma_j^z\,,
\end{eqnarray}
will do. Next, couple the system to a local dephasing bath through
$L_j=\sqrt{\gamma}\,\sigma_j^z$,
so that the Lindblad equation is
\begin{eqnarray}
    \dot\rho = -i[H,\rho] + \gamma\sum_j \left(\sigma_j^z \rho \sigma_j^z - \rho\right)\,,
\end{eqnarray}
or, equivalently, in the Heisenberg picture, an operator $O$ evolves as
\begin{eqnarray}
    \dot O = i[H,O] + \gamma\sum_j \left(\sigma_j^z O \sigma_j^z - O \right)\,.
\end{eqnarray}
Our immediate goal is to understand how the dissipative superoperator
\begin{eqnarray}
    \mathcal D(O) = \gamma\sum_j\left(\sigma_j^z O \sigma_j^z - O \right)\,,
\end{eqnarray}
acts in Krylov space. Toward this end, we expand the operator as $O = \sum_{\alpha} c_\alpha P_\alpha\,,$ in the Pauli-string basis
\begin{eqnarray}
    P_\alpha = \bigotimes_{j=1}^N \sigma_j^{a_j}, \qquad a_j\in\{0,x,y,z\}\,,
\end{eqnarray}
and note that, since $\sigma_j^z$ commutes with $\mathbb I_j\,,$  and $\sigma_j^z$ anticommutes with $\sigma_j^x$ and $\sigma_j^y$, it follows that
\begin{eqnarray}
    \big(\sigma_j^z P_\alpha \sigma_j^z - P_\alpha\big) = \begin{cases} 
        0, & a_j=0,z,\\
        -2P_\alpha, & a_j=x,y\,.
    \end{cases}
\end{eqnarray}
Summing over $j$,
\begin{eqnarray}
    \mathcal D(P_\alpha) = -2\gamma\,N_\perp(P_\alpha)\,P_\alpha\,,
\end{eqnarray}
where $N_\perp(P_\alpha)$ is the number of sites on which the string has $x$ or $y$. The microscopic dephasing superoperator is therefore exactly diagonal in the Pauli-string basis, with eigenvalue proportional to the string’s ``transverse weight". Now take a simple seed, say $O_0=\sigma_0^x$.
Under repeated commutation with a local Hamiltonian, the operator spreads spatially and becomes a superposition of longer Pauli strings. Roughly at Krylov depth $n=0$ it has support on 1 site, and as $n$ grows support extends to more sites and typical strings  acquire more noncommuting $x$ and $y$ content. So along the Krylov sequence $\{|O_n\rangle\}$, the expectation value of $N_\perp$ grows. If we define $\nu_n \equiv \langle O_n|N_\perp|O_n\rangle$, then the projected dissipator matrix elements are
$\langle O_m|\mathcal D|O_n\rangle
=
-2\gamma\,\langle O_m|N_\perp|O_n\rangle$. If the Krylov basis becomes sufficiently typical at large $n$, then $N_\perp$ is approximately diagonal in that basis with $\langle O_m|N_\perp|O_n\rangle
\approx
\nu_n\,\delta_{mn}$. Thus
\begin{eqnarray}
    \mathcal D \approx -2\gamma\sum_n \nu_n\,|n\rangle\langle n|\,,
\end{eqnarray}
and the dissipator becomes an effective diagonal operator in Krylov space. For local chaotic dynamics, Krylov depth $n$ correlates with operator complexity. In spin chains, complexity growth is accompanied by growth of support size, growth of the number of nontrivial Pauli factors, and growth of the number of sites on which the operator fails to commute with the bath coupling. It is therefore natural to expect that $\nu_n \sim c_0 + c_1 n + \cdots$
over the regime where the Krylov basis resolves increasing operator complexity.\\

\noindent
To leading order then, we can take
$\nu_n \approx a n + b$. Then
$\mathcal D_{\rm eff}
\approx
-2\gamma(a\hat n+b)$,
with $\widehat n=\sum_n n\,|n\rangle\langle n|$. Discarding the constant shift $b$, which only gives overall decay, the dissipator is effectively proportional to $\widehat n$ with the corresponding effective jump operator scaling like
$L_{\rm eff}\sim \sqrt{\widehat n}$
or, $L_{\rm eff}\sim \widehat n$, depending on whether we match the dissipator or the jump itself.
For our dephasing toy model, taking $L_{\rm eff}\propto \widehat n$ is the simplest coarse-grained implementation.

\bibliographystyle{JHEP}
\bibliography{biblio.bib}

\end{document}